# Corrugated probe for SNOM – Optimization of energy throughput via plasmon excitation

Tomasz J. Antosiewicz, Tomasz Szoplik

Faculty of Physics, University of Warsaw, Pasteura 7, 02-093, Warsaw, Poland

Correspondence to: Tomasz J. Antosiewicz, Tel: +48-22-55-46855; Fax: +48-22-55-46-882;

e-mail: tantos@igf.fuw.edu.pl

**Abstract**

In a previous paper we proposed a modification of metal-coated tapered-fibre aperture probes for scanning near-field optical microscopes (Antosiewicz & Szoplik, 2007b). The modification consists of radial corrugations of the metal-dielectric interface oriented inward the core. Their purpose is to facilitate the excitation of propagating surface plasmons, which increase the transport of energy beyond the cut-off diameter and radiate a quasi-dipolar field from the probe output rim (Antosiewicz & Szoplik, 2007a). An increase in energy output allows for reduction of the apex diameter, which is the main factor determining the resolution of the microscope. In FDTD simulations we analyse the performance of the new type to SNOM probe. We aim at achieving of maximum energy throughput in probes with corrugations that may be realized in a glass etching process.

**Introduction**

Optical observation of the micro- and nanoworld is a fundamental element of the scientific method and naturally we strive for the best accuracy available. In aperture optics, due to the wave properties of light it is not possible to image objects with accuracy better than half a wavelength. To circumvent this diffraction limit a number of tools had been developed, including the scanning electron microscope, where this limitation is negligible, and the atomic force microscope (AFM), where object surface morphology is retrieved from van der Vaals forces between a probe tip and an object. Another approach, using visible light, is based on an idea of Synge (Synge, 1928). His idea of using a subwavelength diameter beam to inspect an object led to the creation of the scanning near-field optical microscope (SNOM) (Pohl *et al.*, 1984). Such a microscope when combined with a method to control the aperture – sample distance provides information on the optical properties of a sample as well as on its topography (Betzig *et al.*, 1992).

The key parameter determining the resolution achievable by aperture SNOMs is the diameter of the unmetalized probe apex (Ohtsu, 1998; Kim & Song, 2007; Novotny & Hecht, 2007). In the narrow end part of the tip between the cut-off diameter and the aperture the illuminating beam does not propagate any more and there exists an evanescent wave only. Thus a small diameter of the apex reduces the intensity of the transmitted field. Several methods to counter this limitation were proposed, *eg.* large cone angle probes (Lazarev *et al.*, 2003; Haber *et al.*, 2004; Yang *et al.*, 2007), asymmetrically structured probes (Yatsui *et al.*, 1997) or triple tapered probes (Monobe *et al.*, 1998; Yatsui *et al.*, 1998). Large cone angle probes increase transmission by up to two orders of magnitude by virtue of shortening the distance for which evanescent solutions exist. Asymmetrically structured probes enhance the excitation of the $HE_{11}$ mode which has better transmission parameters in metalized fibres than other modes. The triple tapered probes, in turn, combine the characteristics of both previous tips leading up to a high increase in total radiated power. A resolution of about 10 nm can be achieved in measurements with SNOM and AFM using a method called transmission-based SNOM (Hopman *et al.*, 2007).

Recently, another method to increase the near-field intensity in the vicinity of subwavelength apertures in screen by surface plasmons has been proposed (Jin and Xu, 2005; Tanaka *et al.*, 2006). By carefully designing the apertures, favourable conditions for an efficient excitation of plasmons are created. Plasmons generated inside the structure propagate towards the apex and radiate fields characterized by strong intensities and narrow profiles. As usual in plasmon phenomena, the observed enhancement is very sensitive to the geometry of the apertures. The enhancement was reported for structures made in thin screens. Applying these results to SNOM cantilevers would require the fabrication of such structures at their tips, a difficult feat to achieve.

Despite the above mentioned nanotechnological difficulties surface plasmons can be used successfully to increase the light throughput in SNOM probes. In this paper we develop the idea presented in a recent publication (Antosiewicz and Szoplik, 2007b) where a modification to the aperture SNOM probe by introducing corrugations at the interface between the dielectric-core and metal coating was proposed. The role of these corrugations is to enhance the generation of surface plasmons, which propagate beyond the cut-off diameter, on the inside surface of the metal-coating of the tip. The surface plasmons propagate from the corrugations to the aperture rim where they radiate a quasi-dipolar field (Antosiewicz & Szoplik, 2007a). Their generated flux combined with the evanescent field from the aperture clearance create a signal stronger than possible in an uncorrugated SNOM probe. In this paper we analyse the dependence of probe radiating properties on the geometry of introduced corrugations.

**Simulations**

As reference of SNOM probe radiation we choose a probe that tapers down from a waveguide with a diameter of 2 µm to a 50 nm aperture. The taper angle is 20 degrees because we want to analyse how corrugations improve low transmittance and have a tapered fibre long enough to etch grooves. An angle smaller than 20° would result in a long tapered part and would increase the computational area considerably. The core is made of dispersionless silica with refractive index $n_g$ = 1.449 and the cladding is a 70 nm thick layer of silver with dispersion described by the Drude formula

$$\varepsilon(\omega) = \varepsilon_\infty - \omega_p^2 [\omega(\omega + i\Gamma)]^{-1}. \tag{1}$$

We use the following parameters: $\varepsilon_\infty = 3.70$, plasma frequency $\omega_p = 13673$ THz, and damping frequency $\Gamma = 27.35$ THz calculated by Sönnichsen (Sönnichsen, 2001) from experimental data on reflection and transmission of silver films obtained by Johnson and Christy (Johnson and Christy, 1972).

A model of a classical single taper aperture SNOM probe, called in this paper *smooth*, is shown in Fig. 1a. The modified probes with circular and oval corrugations are shown in Figs 1b and 1c, respectively. The first modification shown in Fig. 1b consists of circular indentations of 40 nm radius in the silica core. The centre of the circles is not on the dielectric-metal interface, but is moved radially away from the axis by $\Delta r = r_0 \tan(\alpha)$, where $r_0$ is the corrugation's radius and $\alpha = 20º$ is the taper angle. That is grooves are shallower then semicircular. The second type of corrugations is an oval indentation of 30 nm depth and curvature slightly shifted from the dielectric-metal interface by $\Delta r$. This shift is implemented so that the angles between the sides of the cone and the corrugations are obtuse, which probably would be the case for such fabricated structures.

In our investigations we scan the lattice period $\Lambda$ of six corrugations and the relative shift $s$ of the lattice with respect to the apex, called in this paper *offset*, for both structures. Additionally, in the case of the oval corrugations we change their width. Unless mentioned otherwise, the values of the lattice period and the *offsets* are calculated along the axis of the tip, not along the metal-dielectric interface. In Finite-Difference Time-Domain (FDTD) simulations using the EMFIDES code (Saj, 2005) we analyse the transmission properties of corrugated SNOM probes and compare them with results obtained for a *smooth* tapered tip. Most of the simulations are two-dimensional in order to use a fine computational grid of 0.5 nm space discretization. Full three-dimensional simulations on an 8 times coarser grid are also performed to verify qualitatively the accuracy of the 2D simulations. The excitation signals used in the analysis are a CW Gaussian beam and a broad-band Gaussian impulse of a spatial profile that is 2.5 times smaller than the transverse dimensions of the core. Both excitation signals are polarized linearly with the electric field in the plane of the structure for 2D simulations.

**Optimisation of corrugation parameters**

Important parameters that define the achievable SNOM resolution are energy throughput and the full-width at half-maximum (FWHM) of the output beam which define the optical properties of probes. We compare the results obtained for modified probes with the reference values calculated for a *smooth* SNOM aperture probe.

To illustrate how corrugations influence the propagation of plasmon waves we present time-averaged energy distributions in three analysed probe structures (Fig 2.: *smooth* (a)-(c), circular $\Lambda$ = 445 nm, $s$ = 173 nm (d)-(f) and oval $\Lambda$ = 325 nm, $s$ = 87 nm (g)-(i) corrugations) for three wavelengths: 450 nm, 510 nm and 600 nm in columns from left to right. Structures are chosen such that high and low transmission occurs at the same wavelengths. As reference we show the field distribution in a *smooth* probe in Figs 2a-c. In a *smooth* probe for all three wavelengths the incident wave is reflected at the cut-off diameter plane and does not propagate farther. Here a propagating surface plasmon on the dielectric-metal interface is not excited because its wavevector is larger than that of the wave in the dielectric core and only localised plasmons at the interface are present. Thus the only energy that can be emitted from the apex of a *smooth* probe comes from the evanescent solution in the narrow part of the SNOM tip and is low.

Introduction of corrugations allows for coupling of the wavevector of incident light to that of a travelling plasmon wave via the spatial frequency components of the structure.

Transmission is maximum when the high intensity fields at the dielectric-metal interface are located before the last corrugation. In the structure with circular corrugations (Fig. 2e) the localised plasmon is centrally placed between the first and second indentations and in the structure with oval notches (Fig. 2h) the plasmon is located at the end of the second corrugation. For both structures we observe surface waves travelling beyond the cut-off diameter. These plasmons carry part of the incident energy to the probe apex. This transport mechanism is crucial for the enhancement of energy throughput in corrugated SNOM tips. Low transmission for the λ = 600 nm case for both types of corrugations results from the cut-off diameter located far away from the tip end. Here, light does not couple to travelling plasmons because its wavevector is too short and only a high intensity localised plasmon is generated. For the λ = 450 nm case the cut-off diameter lies closer to the probe end, however, transmission is low, because also for this case travelling surface plasmons are not excited.

Figures 3 and 4 show transmissions of probes with circular and oval corrugations, respectively, calculated for different lattice constants and normalized with respect to the transmission of a *smooth* probe. For both types of corrugations with the increase of the lattice period the position of the maximum enhancement shifts towards longer wavelengths. The redshift connected with the increase of the lattice period can be explained in the following way. Incident light cannot excite a propagating plasmon on a flat surface, because the free space wavevector $k_0$ is shorter than that of a surface wave $k_{sp}$. To overcome this momentum mismatch various methods have been suggested, *e.g.* prism coupling in two configurations or grating coupling (Maier, 2007). The grating coupling method makes use of the reciprocal vector $\gamma$ of the grating $\gamma = 2\pi\Lambda^{-1}$, where $\Lambda$ is the diffraction grating constant, to shift the wavevector of impinging light to the value of $k_{sp}$

$$k_{sp} = k_0 n_g \sin\varphi \pm m\gamma, \qquad (2)$$

where $\varphi$ is the angle of incidence, $n_g$ is the refractive index of glass and *m* is an integer. We recall that for a smooth silver-glass interface the surface plasmon wavevector $k_{sp}$

$$k_{sp}(\omega) = \frac{\omega}{c}\sqrt{\frac{\varepsilon_{Ag}(\omega)\varepsilon_g(\omega)}{\varepsilon_{Ag}(\omega)+\varepsilon_g(\omega)}} \qquad (3)$$

depends on the light wavevector in vacuum $k_0 = \omega/c$ and dielectric functions of both media.

For circular corrugations the transmission enhancement values reach a peak value for the lattice period 180 nm (Fig. 3), while for the oval ones saturate at 345 nm and remain constant within the analysed spectral range (Fig. 4). This saturation of transmission enhancement in the case when the lattice of oval grooves exceeds 345 nm is promising from the practical point of view. Large lattice values should ease the groove etching procedure.

In the second series of simulations a lattice period that gave the best transmission results is chosen and the *offset* is varied. It is calculated as the distance from the probe aperture to the centre of the first curvature of an oval corrugation (right end of oval corrugation in Fig. 1c) The calculated data is shown in Figs 5 and 6 for circular and oval corrugations, respectively. Transmission spectra of the analysed SNOM tips are affected differently by the change of the *offset*. The general tendency for circular corrugations (see Fig. 5) is similar to the trend observed for a changing lattice constant $\Lambda$. The spectral position of the maximum enhancement value shifts towards longer wavelengths with an increasing *offset* and has a peak at $s = 180$ nm. For the oval corrugations, when the *offset* is decreasing, the energy throughput enhancement increases and shifts towards shorter wavelengths. A summary of the data presented in Figs 3-6 is presented in Fig. 7. Here we show the normalized maximum intensities and their spectral positions as a function of the lattice constant $\Lambda$ (Figs 7a and 7c) and the *offset s* (Figs 7b and 7d) for circular (Figs 7a and 7b) and oval (Figs 7c and 7d) corrugations to show the tendencies observed in the above numerical experiments. The highest normalized intensity exceeds that of a *smooth* tip by nearly a factor of 16 (Fig. 7b).

For circular grooves the distinct maximum intensities appear for wavelength 520 nm, $\varLambda$ = 395 nm and $s$ = 180 nm. For oval grooves maximum energy throughput exceeding that of a *smooth* tip by a factor of nearly 10 is observed for the range of $\varLambda$ from 325 to 425 nm and $s$ = 87 nm for various wavelengths.

The other important parameter defining a good SNOM probe is the FWHM of its radiated signal. We measure this value for all analysed cases and compare it with the width of the reference beam for a *smooth* probe. In Figs 8 and 9, for tips with circular and oval corrugations respectively, we present a comparison between the reference FWHM and the corresponding values for selected structures with different lattice constants. We observe that the corrugations do not improve the width of the emitted signal and yield a resolution comparable to a *smooth* probe. Thus, improvement of resolution should come from the decrease of the diameter of the apex while keeping the energy output at detectable levels.

**Fabrication prospects**

Technological means of corrugated probe fabrication have a profound influence on achievable widths of grooves and thus on light throughput enhancement. To our knowledge nanogroove etching in tapered $SiO_2$ fibres was never considered (*e.g.* Ohtsu, 1998; Lazarev *et al.*, 2003; Haber *et al.*, 2004; Yang *et al.*, 2007; Novotny & Hecht, 2007). One may expect that wider grooves are easier to implement. Therefore we vary the oval groove width *d* to assess its influence on the transmission properties. This is analysed in two ways. In the first case the elongation by 10 nm increments affects the near end of the groove and in the second the far end, both with respect to the probe aperture. The results, presented in Figs 8 and 9 respectively, are shown for half of the analysed structures because of clarity. We observe, that in the first case for longer corrugations (Fig. 10) the intensity of radiated energy reaches a gain factor of 23 and the spectral location of the maximum shifts towards shorter wavelengths. An analysis of the Poynting vector distribution (see Fig. 12) shows that the first corrugation from the apex forms what can be called an energy concentrator. Within this small volume surface plasmons have especially high energy and their outward radiation flux is greater than for plasmons bound in larger volumes. A similar tendency of increasing transmission with a decreasing *offset* is also observed when we only change the distance between the lattice and the probe apex and not the groove width. Thus it is possible to increase energy throughput by placing the corrugations close to the aperture of the tip, however this distance is limited by the probe fragility. The second way of increasing the width of the corrugations does not affect the groove-apex distance. Figure 11 shows that an increasing *offset* with a constant groove – apex distance has a negligible impact on the performance of SNOM probes. Both the maximum transmission value and spectral location remain virtually unchanged. This is advantageous for fabrication purposes, as etching relatively wide grooves presents less of a challenge than narrow ones.

**Summary**

This research aims at resolution improvement of SNOMs. Light throughput in corrugated metal-coated tapered fibre probes is calculated using the FDTD method and compared with that of a smooth, single tapered tip. The influence of width and separation of etched grooves on enhanced transmission is investigated for a wide range of parameters. Comparison of high resolution two dimensional (2D) simulations with three dimensional ones confirms the qualitative agreement of both methods. The only difference is the location of the cut-off diameter, which in the 2D case appears closer to the apex. We conclude that circular grooves are more efficient in plasmon coupling than the oval ones, however the latter have two advantageous features. The first is a high enhancement of the output field observed for a wide range of wavelengths and groove widths. The second is connected with etching feasibility.

**Acknowledgements**

This research was sponsored by the Polish grants 37/COS/2006/03 and MNiI 1415/08. We also acknowledge the support from the 6FP Network of Excellence Metamorphose and COST action MP0702.

**References**

Antosiewicz, T.J. & Szoplik, T. (2007a) Description of near- and far-field light emitted from a metal-coated tapered fiber tip. *Opt. Express* **15**, 7845-7852.

http://www.opticsinfobase.org/abstract.cfm?URI=oe-15-12-7845

Antosiewicz, T.J. & Szoplik, T. (2007b) Corrugated metal-coated tapered tip for scanning near-field optical microscope. *Opt. Express* **15**, 10920-10928.

http://www.opticsinfobase.org/abstract.cfm?URI=oe-15-17-10920

Betzig, E., Finn, P.L. & Weiner, J.S. (1992) Combined shear force and near-field scanning optical microscopy. *Appl. Phys. Lett.* **60**, 2484-2486.

Haber, L.H., Schaller, R.D., Johnson, J.C. & Saykally, R.J. (2004) Shape control of near-field probes using dynamic meniscus etching. *J. Microsc.* **214**, 27-35.

Hopman, W.C.L., Stoffer, R. & de Ridder, R.M. (2007) High-Resolution Measurement of Resonant Wave Patterns by Perturbing the Evanescent Field Using a Nanosized Probe in a Transmission Scanning Near-Field Optical Microscopy Configuration. *J. Lightwave Technol.* **25**, 1811-1818. http://www.opticsinfobase.org/abstract.cfm?URI=JLT-25-7-1811

Jin, E.X. & Xu, X. (2005) Obtaining super resolution light spot using surface plasmon assisted sharp ridge nanoaperture. *Appl. Phys. Lett.* **86**, 111106.

Johnson, P. & Christy, R. (1972) Optical Constants of the Noble Metals. *Phys. Rev. B* **6**, 4370-4379

Kim, J. & Song, K.B. (2007) Recent progress of nano-technology with NSOM. *Micron* **38**, 409-426.

Lazarev, A., Fang, N., Luo, Q. & Zhang, X. (2003) Formation of fine near-field scanning optical microscopy tips. Part I. By static and dynamic chemical etching. *Rev. Sci. Instrum.* **74**, 3679-3683.

Maier, S.A. (2007) Plasmonics: Fundamentals and Applications. Springer, New York.

Mononobe, S., Saiki, T., Suzuki, T., Koshihara, S. & Ohtsu, M. (1998) Fabrication of a triple tapered probe for near-field optical spectroscopy in UV region based on selective etching of a multistep index fiber. *Opt. Comm.* **146**, 45-48.

Novotny, L. & Hecht, B. (2007) Principles of Nano-Optics. Cambridge University Press, Cambridge.

Ohtsu, M. (1998) Near-Field Nano/Atom Optics and Technology. Springer, Tokyo.

Pohl, D.W., Denk, W. & Lanz, M. (1984) Optical stethoscopy: Image recording with resolution l/20. *Appl. Phys. Lett.* **44**, 651-653.

Saj, W. (2005) FDTD simulations of 2D plasmon waveguide on silver nanorods in hexagonal lattice. *Opt. Express* **13**, 4818-4827.

http://www.opticsinfobase.org/abstract.cfm?URI=oe-13-13-4818

Sönnichsen, C. (2001) Plasmons in metal nanostructures, PhD Thesis Ludwig-Maximilians-Universtät München, München.

Stöckle, R.M., Fokas, C., Deckert, V., Zenobi, R., Sick, B., Hecht, B. & Wild, U.P. (1990) High-quality near-field optical probes by tube etching. *Appl. Phys. Lett.* **75**, 160-162.

Synge, E.H. (1928) A suggested method for extending the microscopic resolution into the ultramicroscopic region. *Phil. Mag*. **6**, 356-362.

Tanaka, K., Tanaka, M. & Sugiyama, T. (2006) Creation of strongly localized and strongly enhanced optical near-field on metallic probe-tip with surface plasmon polaritons. *Opt. Express* **14**, 832-846. http://www.opticsinfobase.org/abstract.cfm?URI=oe-14-2-832

Yatsui, T., Kourogi, M. & Ohtsu, M. (1998) Increasing throughput of a near-field optical fiber probe over 1000 times by the use of a triple-tapered structure. *Appl. Phys. Lett.* **73**, 2090-2092.

Yang, J., Zhang, J., Li, Z. & Gong, Q. (2007) Fabrication of high-quality SNOM probes by pre-treating the fibres before chemical etching. *J. Microsc.* **228**, 40-44.

Yatsui, T., Kourogi, M. & Ohtsu, M. (1997) Highly efficient excitation of optical near-field on an apertured fiber probe with an asymmetric structure. *Appl. Phys. Lett.* **71**, 1756-1758.

**Figure captions**

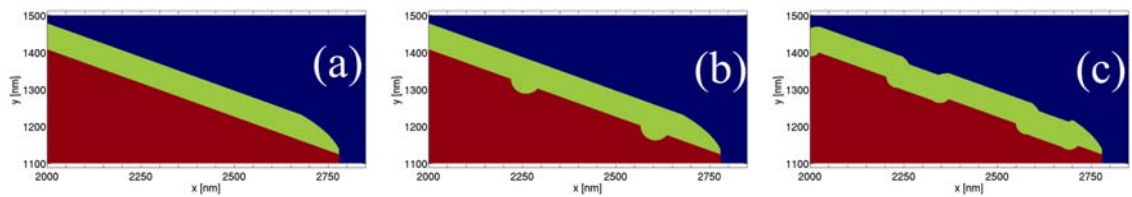

Fig. 1. Modeled tip structures (a) *smooth* tip, (b) tip with circular grooves, (c) tip with oval grooves. Colours indicate: glass core – dark red, metal coating – green, vacuum – blue. The pictures show, because of clarity, only the symmetrically cut, narrow end of the tips.

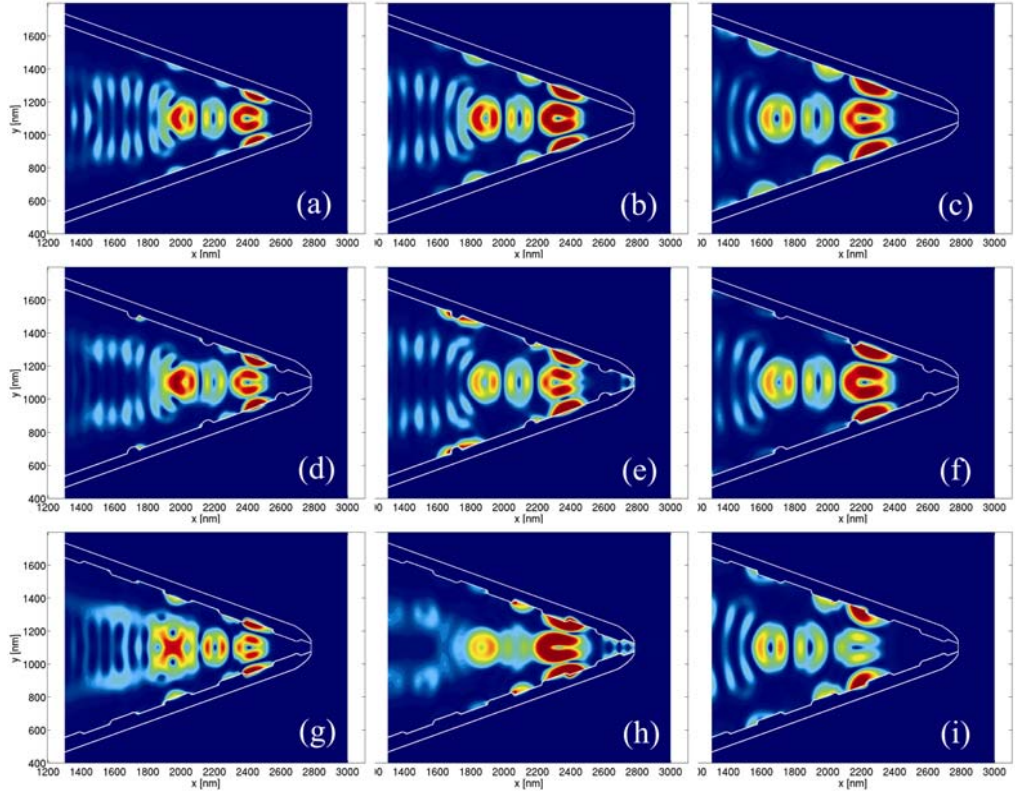

Fig. 2. Time-averaged energy distributions in three analysed structures: *smooth* probe (a)-(c), probe with circular grooves of $\Lambda = 445$ nm and $s = 173$ nm (d)-(f) and oval corrugations of $\Lambda = 325$ nm and $s = 87$ nm (g)-(i). Wavelengths shown: 450 nm, 510 nm and 600 nm in left, center and right columns respectively.

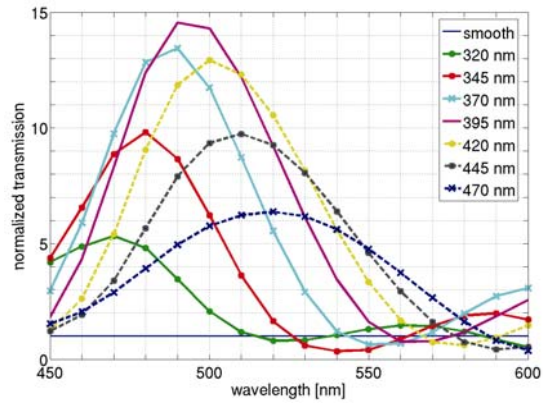

Fig. 3. Transmission for SNOM probes with circular corrugations with different lattice constants calculated for the spectral range 450-600 nm. Inlet gives lattice constant $\Lambda$ values. For each wavelength values are normalized to transmission of a *smooth* tip.

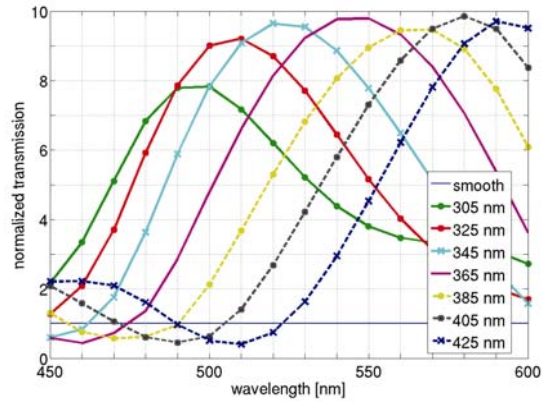

Fig. 4. Transmission for SNOM probes with oval corrugations with different lattice constants calculated for the spectral range 450-600 nm. Inlet gives lattice constant $\Lambda$ values. For each wavelength values are normalized to transmission of a *smooth* tip.

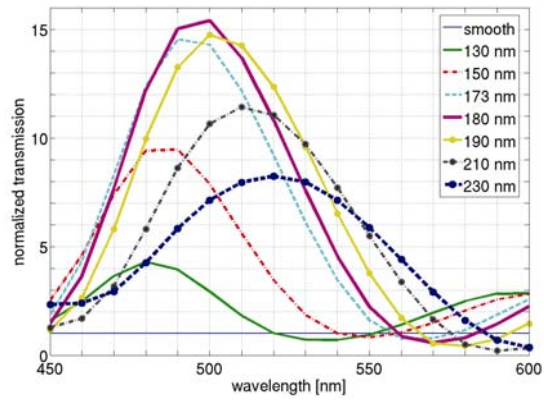

Fig. 5. Transmission for SNOM probes with circular corrugations with different *offset* values shown in the inlet calculated for the spectral range 450-600 nm. For each wavelength values are normalized to transmission of a *smooth* tip.

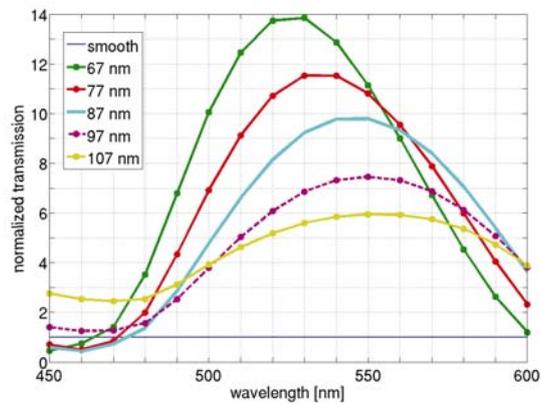

Fig. 6. Transmission for SNOM probes with oval corrugations with different *offset* values shown in the inlet calculated for the spectral range 450-600 nm. For each wavelength values are normalized to transmission of a *smooth* tip.

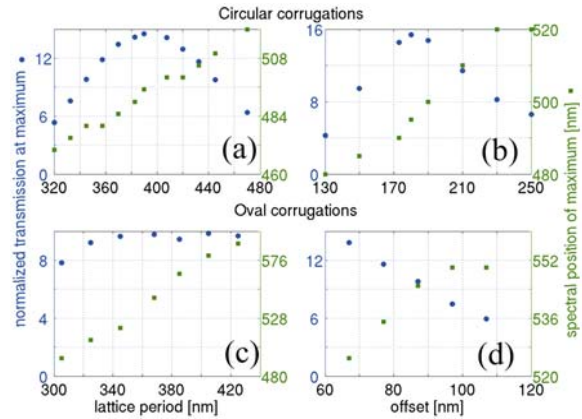

Fig. 7. Normalized transmissions at maximum (circles, left y-axis) and the spectral position of the maximum (squares, right y-axis) calculated for circular (a), (b) and oval (c), (d) corrugations as a function of lattice constant (a), (c) and *offset* (b), (d). In plots (a) and (c) the *offsets* are constant and equal 173 nm and 200 nm respectively; in plots (b) and (d) the lattice periods are constant and equal 395 nm and 365 nm respectively. These plots have data points which were omitted in Figs 3-6 for clarity.

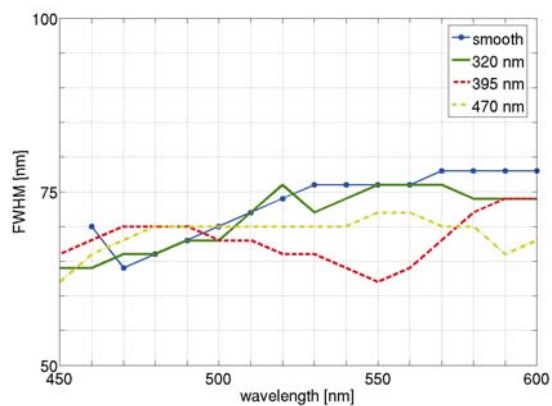

Fig. 8. Comparison of FWHM for a *smooth* probe and probes with circular corrugations of different lattice constants.

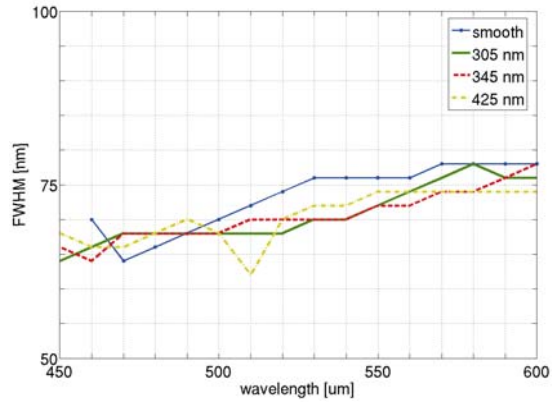

Fig. 9. Comparison of FWHM for a *smooth* probe and probes with oval corrugations of different lattice constants.

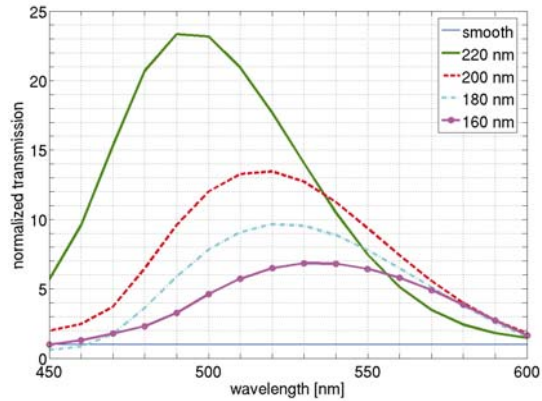

Fig. 10. Normalized transmission for SNOM probes with oval corrugations of varied lengths. With respect to Fig. 1c the position of the left side of the corrugations is kept constant and the right is varied, *i.e.* the grooves are longer and approach the probe end. Plots are labelled by the distance of groove edges.

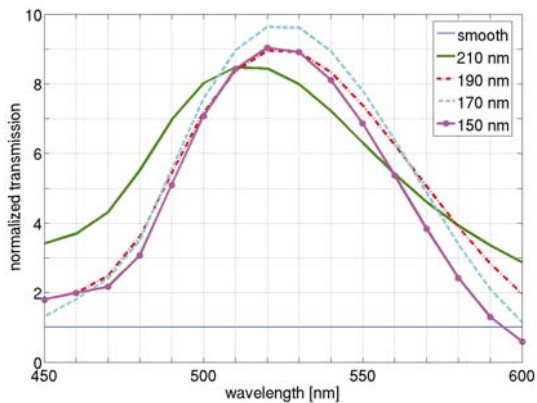

Fig. 11. Normalized transmission for SNOM probes with oval corrugations of varied lengths. With respect to Fig. 1c the position of the right side of the corrugations is kept constant and the left is varied, *i.e.* the grooves are longer or shorter, but begin at the same place with respect to the probe aperture. Plots are labelled by the distance of groove edges.

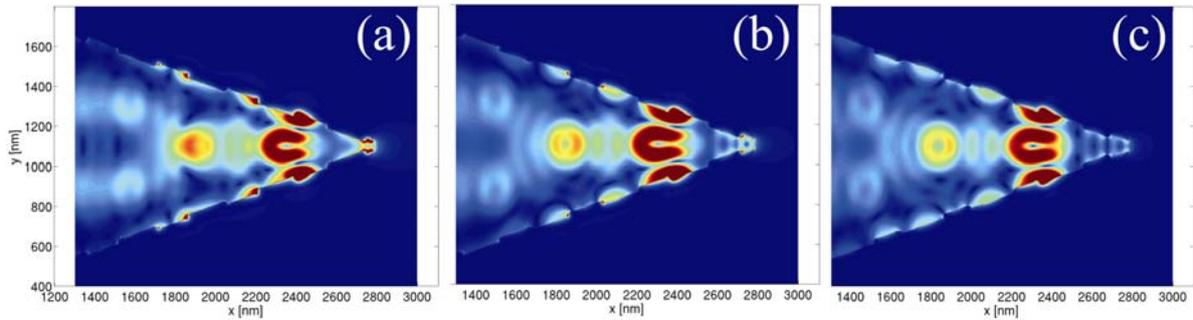

Fig. 12. Time-averaged energy distributions in structures with oval corrugations of different length: in (a) 220 nm, (b) 190 nm, (c) 160 nm. For the wavelengths chosen (a) 490 nm, (b) 520 nm, (c) 530 nm transmission is maximum for the given structures. Intensity scales are the same for all subfigures.